\documentclass[11pt,a4paper]{article}
\usepackage{amsmath}
\usepackage[psamsfonts]{amssymb}
\usepackage{amsmath}
\usepackage{epsfig}
\usepackage{graphicx}

\title{On thermodynamics second law in the modified Gauss Bonnet gravity}
\author{H. Mohseni Sadjadi$^1$\footnote{mohsenisad@ut.ac.ir},
\\$^1${\small Department of physics, University of Tehran,}
\\{\small P.O.B. 14395-547, Tehran, Iran}.}

\begin{document}
\maketitle
\begin{abstract}
The second law and the generalized second law of thermodynamics in
cosmology in the framework of the modified Gauss-Bonnet theory of
gravity are investigated. The conditions upon which these laws
hold are derived and discussed.
\end{abstract}

\section{Introduction}

Recently, extended theories of gravity have attracted more
attentions, because besides their possible ability to describe the
inflation in the early universe \cite{early}, they can be proposed
as candidates to explain the present acceleration of the universe
\cite{ac} without requirement to introduce exotic matter with
negative pressure, dubbed as dark energy. A method to generalize
the Einstein theory of gravity is to consider a Lagrangian which,
in addition to the Ricci scalar, includes a general function of
the Gauss-Bonnet invariant\cite{GB}. This generalization has root
in low energy effective action in string theory \cite{root}.

However these modifications of Einstein theory of gravity must be
in agreement with astrophysical data. Severe instabilities may be
appeared in the modified Gauss Bonnet gravity. Quintessence model
coupled to Gauss Bonnet invariant and its stability were studied
in \cite{Tomi}. Also, the constraints that the stability of
perturbation growths puts on modified gravity were considered in
\cite{Baoj}. The investigation of such cosmological perturbations
in the presence of a barotropic fluid can also be found in
\cite{antonio}.

In analogy with black hole thermodynamics, investigating
thermodynamics laws for cosmological horizons, which are present
in many of cosmological models,  has also been the subject of many
studies \cite{therm}. One of these laws is the generalized second
law (GSL) which asserts that the sum of the horizon entropy and
the entropy of the matter field , i.e. the total entropy, is
non-decreasing \cite{gsl}. Our aim is to study the constraints
that this law puts on the functions introduced in modified
Gauss-Bonnet theory of gravity which may be similar to what was
obtained in papers like \cite{antonio} with a different method.

The scheme of the paper is as follows: We consider a spatially
flat Friedman-Lemaitre- Robertson-Walker (FLRW) space-time whose
the future event horizon is in thermal equilibrium with the matter
in its environment. In the framework of the modified Gauss-Bonnet
gravity, by using modified Friedman equations, and by considering
the entropy evolution of dynamical future event horizon (obtained
via the Noether charge method) we study the necessary conditions
for validity of GSL. Via an example, we will show that this law
may not hold in general. The same also occurs in some stages in
phantom dominated universe in the Einstein theory of gravity as
was reported in \cite{pav}.

We use units with $\hbar=c=G_N=k_B=1$ and the metric signature is
$(-,+,+,+)$.
\section{Thermodynamics second law and GSL in the modified Gauss-Bonnet gravity}
The action of the modified Gauss-Bonnet gravity is given by
\begin{equation}\label{1}
S=\int d^4x \sqrt{-g}\left({R+f(G)\over
16\pi}+\mathcal{L}_{m}\right),
\end{equation}
where $f(G)$ is a function of the Gauss-Bonnet invariant,
$G=R^2-4R_{\mu \nu}R^{\mu \nu}+R_{\mu \nu \rho \sigma}R^{\mu \nu
\rho \sigma}$, and $R$ is the Ricci scalar curvature.
$\mathcal{L}_{m}$ is the matter Lagrangian density. By variation
of the action with respect to the metric components $g_{\mu \nu}$,
we obtain the field equations
\begin{eqnarray}\label{2}
&&8\pi T^{\mu \nu}=-{1\over 2}g^{\mu\nu}(R+f(G)) +R^{\mu\nu}
+2g^{\mu \nu} R\nabla^2 f_G\nonumber \\
&&-4R^{\mu \nu}\nabla^2 f_G -2R(\nabla^{\mu}\nabla^{\nu})f_G
+4R^{\mu \alpha}\nabla_{\alpha}\nabla^{\nu}f_G+4R^{\nu
\alpha}\nabla_{\alpha}\nabla^{\mu}f_G
\nonumber \\
&&-4g^{\mu \nu}R^{\alpha \beta}\nabla_{\alpha}\nabla_{\beta}f_G
+4R^{\mu \alpha\nu\beta}\nabla_{\alpha}\nabla_{\beta}f_G+2RR^{\mu
\nu}f_G-4R^{\mu}_{\alpha}R^{\nu \beta}f_G
\nonumber\\
&&+2R^{\mu \alpha\beta\gamma}R^{\nu}_{\alpha\beta\gamma}f_G
+4R_{\alpha \beta}R^{\mu\alpha\beta\nu}f_G,
\end{eqnarray}
where $f_{G}={df\over dG}$ and $T^{\mu \nu}$ are the matter
energy-momentum tensor components.

We consider a spatially flat FLRW space time, endowed with the
metric (in the Cartesian comoving coordinates):
\begin{equation}\label{3}
ds^2=-dt^2+a^2(t)(dx^2+dy^2+dz^2).
\end{equation}
In terms of the Hubble parameter, $H$, the Ricci scalar and
Gauss-Bonnet invariant are given by
\begin{equation}\label{4}
R=6(\dot{H}+2H^2),\,\,\,G=24H^2(\dot{H}+H^2).
\end{equation}
The over dot indicates derivative with respect to $t$.
Eqs.(\ref{2}) give the following modified Friedman equations:
\begin{eqnarray}\label{5}
\dot{H}&=&-4\pi(P_m+\rho_m)+2H^3\dot{f_G}-4H\dot{H}\dot{f_G}-2H^2\ddot{f_G}\nonumber\\
H^2&=&{8\pi\over 3}\rho+{1\over 6}(-f+Gf_G-24H^3\dot{f_G}).
\end{eqnarray}
$\rho_m$ and $P_m$ are energy density and pressure of the matter
component which is assumed to behave as a perfect fluid at large
scale. In this paper we assume that the matter is ordinary with
positive pressure: $P_m>0$.

By setting $f=0$, the above equations reduce to the Friedman
equations in the Einstein theory of gravity. The matter satisfies
the continuity equation:
\begin{equation}\label{6}
\dot{\rho_m}+3H(P_m+\rho_m)=0.
\end{equation}

We assume that the universe possesses a future event horizon whose
radius is given by
\begin{equation}\label{7}
R_h(t):=a(t)\int_t^{\infty}{dt'\over a(t')}\in \Re.
\end{equation}
If there is a big rip at $t_s$, we must replace $\infty$ by $t_s$.
The evolution of the horizon is given by
\begin{equation}\label{8}
\dot{R_h}=HR_h-1.
\end{equation}

The entropy of the dynamical horizon, $S_h$, can be determined by
the Noether charge method\cite{wald} which results in
\begin{equation}\label{9}
S_h=-{1\over 8}\int_{Horizon}\left({\partial R\over \partial
R_{\alpha \beta \gamma \rho}}+f_G{\partial G\over \partial
R_{\alpha \beta \gamma \rho}}\right)\varepsilon_{\alpha
\beta}\varepsilon_{\gamma \rho}dA_{h},
\end{equation}
where $\varepsilon_{\mu \nu}$ are binormal vectors to the horizon
surface and $dA_h$ is the differential surface element. By
applying this result for the future event horizon we obtain:
\begin{equation}\label{10}
S_h={A_h\over 4}\left(1+{4\over R_h^2}f_G\right),
\end{equation}
where $A_h=4\pi R_h^2$ is the area of the future event horizon.
For a de Sitter space-time, $R_h^2={12\over R}$ and,(\ref{10})
becomes
\begin{equation}\label{11}
S_h={A_h\over 4}\left(1+{R\over 3}f_G\right).
\end{equation}
Although linear terms of Gauss-Bonnet invariant in $f$ change the
horizon entropy, but have no influence on $\dot{S_h}$:
\begin{equation}\label{12}
\dot{S_h}=2\pi R_h(HR_h-1)+4\pi \dot{f_G}.
\end{equation}
For a de Sitter space time $R_h={1\over H}$ and $\dot{S_h}=0$.

In a super-accelerated universe (i.e. $\dot{H}>0$), $\dot{R_h}<0$
\cite{sad1}, which in Einstein theory of gravity implies
$\dot{S_h}=2\pi R_h\dot{R_h}<0$. But depending on the form of $f$,
this may not be the case in the modified Gauss-Bonnet theory of
gravity. To see this, let us consider a quasi de Sitter space time
specified by the Hubble parameter
\begin{equation}\label{13}
H=H_0+H_0^2 \epsilon(t)t+\mathcal{O}(\epsilon^2)
\end{equation}
where $\epsilon:={\dot{H}\over H^2}$, $\mid\epsilon\mid\ll 1$, and
$\dot{\epsilon}\sim\mathcal{O}(\epsilon^2)$. We can expand $R_h$
and $G$ around de Sitter point ($H_0$), up to order $\epsilon^2$,
as
\begin{eqnarray}\label{14}
&&G=24H_0^4(1+\epsilon)+96H_0^5t\epsilon,\,\,\dot{G}=96H_0^5\epsilon
,
\nonumber  \\
&&R_h\approx {1\over H}(1-\epsilon),\,\,\,\, \dot{R_h}\approx
-\epsilon.
\end{eqnarray}
We have also
\begin{equation}\label{15}
\dot{f_G}=(f_{GG}(H_0)+f_{GGG}(H_0)(G-G_0))\dot{G}(H_0),
\end{equation}
where $G_0=24H_0^2$. Note that for a given model by a specific
$f(G)$, $H_0$ is the solution of the equation
\begin{equation}\label{16}
G_0f_G(G_0)-f(G_0)=6H_0^2-16\pi \rho.
\end{equation}
Collecting all together the second law of thermodynamics for the
horizon $(\dot{S_h}\geq 0)$ requires:
 \begin{equation}\label{17}
(-1+192H_0^6f_{GG})\epsilon+\mathcal{O}(\epsilon^2)\geq 0.
\end{equation}
If the system is in quintessence (phantom) phase, i.e. $\dot{H}<
(>)0$, we have $\epsilon< (>)0$ and (\ref{17}) leads to
$\left(-1+192H_0^6f_{GG}\right)\leq (\geq)0$.

The stability of modified Gauss-Bonnet gravity around de Sitter
point, $H_0$, and in the vacuum (absence of matter) was
investigated in \cite{viable}.  There was obtained that for
\begin{equation}\label{18}
0<H_0^6f_{GG}(H_0)<{1\over 192},
\end{equation}
the theory is stable. So for vacuum solutions and for stable
models which are perturbed around de Sitter point, the
thermodynamics second law for the future event horizon holds
whenever $\epsilon<0$. For $\epsilon>0$ which leads to $\dot{H}>0$
(super accelerated universe), we obtain $\dot{S_h}<0$.

To investigate the thermodynamics second law for the universe,
besides the horizon entropy, we must also take into account the
entropy of the matter. So let us study time evolution of entropy
of the matter inside the future event horizon (denoted by
$S_{in}$). From the first law of thermodynamics:
\begin{equation}\label{19}
dE=TdS_{in}-PdV,
\end{equation}
we obtain  $\dot{S_{in}}={P_m+\rho_m\over
T}\left(-3HV+\dot{V}\right)$. Then by using (\ref{8}) we obtain
\begin{equation}\label{20}
\dot{S_{in}}=-{A_h\over T}(P_m+\rho_m).
\end{equation}
For the ordinary matter with positive pressure, we have
$\dot{S_{in}}< 0$. The matter is in thermal equilibrium with the
horizon, to which in analogy with the black hole thermodynamics
and independent of the gravity theory resulting the considered
geometry \cite{geom}, the temperature $T={1\over 2\pi R_h}$ is
attributed \cite{dav}.

By using (\ref{5}) we arrive at
\begin{equation}\label{21}
\dot{S_{in}}=\pi R_h^3(2\dot{H}-4H^3\dot{f_G}+8H\dot{H}\dot{f_G}
+4H^2\ddot{f_G}).
\end{equation}
The generalized second law states that the total entropy of the
universe is a non-decreasing function of time. So validity of this
law requires
\begin{equation}\label{22}
\dot{S}_{tot}=\dot{S}_{in}+\dot{S}_h\geq 0.
\end{equation}
Note that as $\dot{S_{in}}<0$, this law requires $\dot{S_h}>0$.

If the matter ingredient is a barotropic matter $P_m=w_m\rho_m$,
(\ref{21}) may be written in a more simple form. Using the
continuity equation
\begin{equation}\label{23}
\rho=\tilde{\rho_0}a^{-3\gamma},
\end{equation}
we find out that the generalized second law is respected whenever
\begin{equation}\label{24}
2\pi R_h(HR_h-1)+4\pi
\dot{f_G}-8\pi^2R_h^3\gamma\tilde{\rho_0}a^{-3\gamma}\geq 0.
\end{equation}
We have defined $\gamma=w_m+1\geq 1$.

In the Einstein theory of gravity, as we have expressed before,
for a (super) accelerated expansion $\dot{S_h}(<)>0$ holds, but we
have also $\gamma(<)>0$ (indicating that the universe is dominated
by (phantom) quintessence dark energy), leading to $S_{in}(>)<0$.
Therefore the generalized second law can be still valid. In our
model both the horizon entropy and Friedman equations are modified
so we may have an accelerated universe even with ordinary matter
with decreasing entropy. Besides, the horizon entropy is not
necessarily decreasing in phantom ($\dot{H}>0$) phase or
increasing in quintessence phase ($\dot{H}<0$) of acceleration. In
a super accelerated universe with an ordinary matter as
$\dot{R_H}<0$ and $\dot{S_{in}}<0$, a necessary conditions
required for GSL to hold is $\dot{f}_G=f_{GG}\dot{G}>0$. Note that
without the correction term in the horizon entropy, GSL would be
violated.

As an illustration of our results, in the following, we will try
to examine the validity of the GSL trough an example. Consider a
super accelerated universe with a big rip singularity at $t=t_s$:
\begin{equation}\label{25}
a={a_0\over (t_s-t)^n},\,\,\ 0<n(\neq 1),
\end{equation}
and dominated by a barotropic perfect fluid $P_m=w_m\rho_m$. The
Hubble parameter is given by $H={n\over (t_s-t)}$. The continuity
equation yields
\begin{equation}\label{26}
\rho=\rho_0(t_s-t)^{3n\gamma},
\end{equation}
where $\rho_0=\tilde{\rho}_0a_0^{-3\gamma}$. From
$G={24n^3(n+1)\over (t_s-t)^4}$ we can write
\begin{equation}\label{27}
(t_s-t)^{-1}=\left({G\over 24n^3(n+1)}\right)^{1\over 4}.
\end{equation}
The above equation may be used to write the second equation in
(\ref{5}) as a differential equation for $f$ in terms of $G$:
\begin{eqnarray}\label{28}
&&-{4\over n+1}G^2f_{GG}+Gf_G-f-6\left({n\over
24(n+1)}\right)^{1\over 2}G^{1\over 2}
\nonumber \\
&&+D\left(24n^3(n+1)\right)^{{3n\over 4}\gamma}G^{-{3n\over
4}\gamma}=0.
\end{eqnarray}
We have defined $D=16\pi \rho_0$. The solution of this
differential equation is
\begin{eqnarray}\label{29}
&&f(G)=C'G+CG^{n+1\over 4}+{(6n(n+1))^{1\over 2}\over 1-n}G^{1\over 2}\nonumber\\
&&+ {4D(n+1)\left(24n^3(n+1)\right)^{{3\over 4}n\gamma}\over
\left(3n\gamma+4\right)\left(3n\gamma+n+1\right)}G^{-{3n\over
4}\gamma},
\end{eqnarray}
where $C'$ and $C$ are two constants. The first term which is
linear in $G$, although appears in the horizon entropy, but has
influence neither on $\dot{S}$ nor on the equation (\ref{5}), so
we put $C'=0$. For $f(G)=0$ the model is the Einstein theory of
gravity, where $H^2={8\pi\over 3}\rho_m$ and the continuity
equation necessitates
\begin{equation}\label{30}
D=6n^2,\,\,\,n=-{2\over 3\gamma}.
\end{equation}
By choosing $C=0$, we get a solution which by applying (\ref{30}),
reduces to the Einstein theory of gravity, i.e. $f=0$. This is
similar to the solution chosen in \cite{root}.

The future event horizon is given by
\begin{equation}\label{31}
R_h={{t_s-t}\over{1+n}}.
\end{equation}
As stated before, $\dot{R_h}<0$. The evolution of the horizon
entropy is described by
\begin{equation}\label{32}
\dot{S_h}=-2\pi{(24n^3(n+1))^{1/4}\over
(n+1)^2}G^{-1/4}+{16\pi\over(24n^3(n+1))^{1/4}}G^{5/4}f_{GG}.
\end{equation}
Without the second term we have always $\dot{S_h}<0$.  By putting
(\ref{29}) (with $C=C'=0$)in(\ref{32}) we get
\begin{eqnarray}\label{33}
\dot{S_h}&=&{12\pi D\gamma n(n+1)(24n^3(n+1))^{{1\over
4}(3n\gamma-1)}\over 3n\gamma+n+1}G^{-{3\over
4}(n\gamma+1)}\nonumber \\
&+& 2\pi \left({24\over n(1+n)^7}\right)^{1\over 4}{3n+1\over n-1}
G^{-{1\over 4}}.
\end{eqnarray}

For $n>1$, thermodynamics second law for the horizon is always
respected. For $n<1$, we may have $\dot{S_h}<0$  when $t$ is
enough far from $t_s$. In this case the main contribution in
$\dot{S_h}$ comes from the term including $G^{-{1\over 4}}$ (note
that $\gamma\geq 1$).

The matter entropy, is obtained as
\begin{equation}\label{34}
\dot{S_{in}}=-{8\pi^2 \gamma \rho_0}{(24n^3(n+1))^{{3\over
4}(1+n\gamma)}\over (1+n)^3}G^{-{3\over 4}(1+n\gamma)}.
\end{equation}
Note that as we have stated before, for ordinary matter
$(\gamma>1)$, $\dot{S_{in}}<0$. So by considering (\ref{32}), we
conclude that a necessary condition for GSL to be respected is
$f_{GG}>0$. The appearance of the explicit form of $f$ in
$\dot{S_{tot}}$, is due to the entropy modification in modified
Gauss-Bonnet gravity (see (\ref{10})). By using (\ref{33}) and
(\ref{34}), $\dot{S_{tot}}$ may be obtained in terms of $G$. In
this way $\dot{S_{tot}}\geq 0$ implies that only for times
satisfying
\begin{equation}\label{35}
BG^{-{3\over 4}n\gamma-{1\over 2}}\geq -A,
\end{equation}
where
\begin{eqnarray}\label{36}
&&A=\left({24\over n(1+n)^7}\right)^{1\over 4}{3n+1\over n-1},\nonumber \\
&&B=\ddot{6Dn\gamma \left({-3\gamma n^3 + 2n^2+3n+1\over
(1+n)^2(3n\gamma+n+1)}\right)}\left(24n^3(n+1)\right)^{{3\over
4}n\gamma-{1\over 4}},
\end{eqnarray}
GSL is valid.

Note that $A\neq 0$, therefore  an adiabatic expansion is
forbidden in this model. We have $A>0$ for $n>1$, and as in our
model $\gamma\geq 1$, for $B>0$ to be hold it is necessary to have
$n<1.488$. It is clear that for the case: $A>0$,\,\,$B>0$, GSL is
respected in all times. In other cases GSL may not be valid, e.g.
if $n<1$, and in the limit $t\to t_s$ (near the big rip), GSL is
not respected, this may due to quantum effects near the big rip
ignored in our classical computation. Also for $t_s\gg t$ such
that $DG^{-{3\over 4}n\gamma-{1\over 2}}\ll 1$ (or
$G_NDG^{-{3\over 4}n\gamma-{1\over 2}}\ll 1$ in units with
$\hbar=c=1, G_N\neq1$), GSL is clearly violated for $n>1.488$ and
$\gamma\geq 1$.

\section{Summary}

In this paper we considered spatially flat FLRW universe in the
framework of the modified Gauss Bonnet gravity (see (\ref{1})). We
assumed that the universe has a future event horizon (see
(\ref{7})). Using the modified Friedman equations (see (\ref{5})),
and the Wald entropy for dynamical horizons (see (\ref{9})),
thermodynamics second law for the horizon, as well as for the
matter component, and finally GSL for the universe were
investigated.  The conditions required for validity of these laws
were obtained (see (\ref{12}) and (\ref{24})), and through an
example by obtaining an exact solution for modified Gauss Bonnet
cosmology ( see \ref{29}) in a super accelerated universe, we
clarified our results.

\end{document}